\documentstyle[preprint,revtex]{aps}
\tightenlines
\begin{document}
\draft
\hyphenation{Nijmegen}
\hyphenation{Weid-mann}
\hyphenation{Stich-ting Fun-da-men-teel On-der-zoek Ma-te-rie}
\hyphenation{Ne-der-land-se Or-ga-ni-sa-tie We-ten-schap-pe-lijk}

\begin{title}
{\bf On the pion-nucleon coupling constant}
\end{title}

\author{Vincent Stoks, Rob Timmermans,\cite{postdoc}
                          and J.J. de Swart\cite{email}}

\begin{instit}
Institute for Theoretical Physics, University of Nijmegen,
Nijmegen, The Netherlands
\end{instit}

\receipt{}

\begin{abstract}
In view of the persisting misunderstandings about the
determination of the pion-nucleon coupling constants in the
Nijmegen multienergy partial-wave analyses of $pp$, $np$, and
$\overline{p}p$ scattering data, we present additional
information which may clarify several points of discussion.
We comment on several recent papers addressing the issue of the
pion-nucleon coupling constant and criticizing the Nijmegen
analyses.
\end{abstract}
\narrowtext

\nonum\subsection{\bf I. INTRODUCTION}
There appear to exist some misunderstandings concerning
the determination by our group of the pion-nucleon
coupling constants in multienergy partial-wave analyses
of $pp$~\cite{Ber90}, combined $pp$ and $np$~\cite{Klo91},
and $\overline{p}p$~\cite{Tim91} scattering data.
We have summarized the recent determinations of
the pion-nucleon coupling constants in Table~\ref{tab:pion}.

In their latest analysis of $\pi^{\pm}p$ scattering data
the VPI\&SU group has obtained a value for the pion-nucleon
coupling constant consistent with our values~\cite{Arn90}.
Since the values $f^2_{N\!N\pi}$ $\simeq$ 0.075 found by us
(see also Refs.~\cite{Rom86,Ber87,Van89,Sha90,Elb91})
are consistently lower than the value $f^2_{N\!N\pi}$ = 0.079
found in the Karlsruhe-Helsinki analyses of $\pi^{\pm}p$ scattering
data~\cite{Hoh75,Koc80,Dum83}, a number of
papers~\cite{Tho89,Gro90,Hol90,Coo90,Mac91,Eri92,Hai92} have been
published commenting on our results. In several of these papers
one attempts to find an explanation for the difference by
trying to point out alleged shortcomings and so-called systematic
errors pertaining to our method of analysis.
In view of the existing confusion,
we have looked again into these and other matters and carefully
reexamined possible sources of systematic errors. We felt
it would be useful to supply some additional information to
further clarify these issues and to address a few points where
some of these papers criticizing us are in error.

A very important point that has not been
recognized and appreciated enough is
that the basis of our accurate determination of the pion-nucleon
coupling constants and of the nucleon-nucleon phase shifts is
our {\bf multienergy} partial-wave analysis of all nucleon-nucleon
scattering data below $T_{\rm lab}$ = 350 MeV.
This multienergy partial-wave analysis is much more
sophisticated and computer intensive than
other similar analyses of nucleon-nucleon scattering
data. The energy dependence of the phase shifts in our analysis
is described in a much better way. As a consequence,
we end up with a multienergy partial-wave solution which gives
an excellent fit to all nucleon-nucleon scattering data
below $T_{\rm lab}$ = 350 MeV. The phase shifts and pion-nucleon
coupling constants as determined by this multienergy partial-wave
analysis are the best possible values for these quantities
that can be obtained from these data.
We strongly feel that statements about the
pion-nucleon coupling constant made with the help of
single-energy analyses or, even worse, analyses of individual
experiments with the help of potential models cannot, in any way,
be compared in quality with the results of these multienergy
analyses. The reason is that in such analyses at one particular
energy the information about the energy dependence of the
phase shifts due to one-pion exchange cannot be incorporated.
In a multienergy analysis the complete database is used
and energy-dependent constraints are properly accounted for.
This results in a much better and more precise value for the
coupling constant, with a minimal model dependence. For a proper
determination the whole database is needed and arguments based
on studies of specific experiments are not so reliable.

In our discussion, we take mostly the Nijmegen
multienergy analysis of all $pp$ scattering data below $T_{\rm lab}$
= 350 MeV~\cite{Ber90} as specific example. The value for the
$pp\pi^0$ coupling constant obtained in this analysis is
undoubtedly the most compelling evidence for
a low pion-nucleon coupling constant, since in this case
the required theoretical input is rather small and
practically model independent. Therefore, the possible
systematic error in this case is very small.
Actually, the statistical error on the $np\pi^{\pm}$ coupling
constant obtained in the $np$ analysis is smaller than
the statistical error on the $pp\pi^0$ coupling constant,
although the $np$ data are less accurate and less varied than
the $pp$ data. The reason, we think, is that in $np$ scattering
pion exchange can be probed more easily, since in
$pp$ scattering one always has to deal with the infinite-range
repulsive Coulomb potential and one first has to strip the
$pp$ data from this long-range Coulomb contribution in order
to get a handle on pion exchange.
Nevertheless, in the $pp$ analysis we feel almost certain that
the systematic error on the $pp\pi^0$ coupling constant is small,
since in this case we explicitly investigated all thinkable
sources of systematic errors and found no significant effects.
We stress that our emphasis on the $pp\pi^0$ coupling constant
does not mean that we have doubts about the correctness of
the determination of the pion-nucleon coupling constants
in the $np$ and $\overline{p}p$ partial-wave analyses.
It simply is more difficult to do a similar thorough
study for the $np$ and $\overline{p}p$ cases, where the amount
of theoretical input into the analyses is larger.
For instance, in these cases it is necessary to make from the
start some theoretical assumptions about the validity of
charge independence.

In the following sections of this paper
we will report on our search for systematic errors
in the value for the $pp\pi^0$ coupling constant in the $pp$
analysis. Part of the reason we have been able to do this
comprehensive investigation is that recently a lot
of computing power has become available to us. We begin
the next section by discussing some statistics relevant
to the analyses. Then we will
subsequently discuss the influence of form-factor effects,
the sensitivity of the different types of observables
to the pion-nucleon coupling constant, and its determination
from individual partial waves and from data in different
energy ranges. Next, we will investigate
more closely some particular $pp$ and $np$ scattering
experiments that are brought up in connection with the
determination of the coupling constant. Finally,
we spend a few words on the Nijmegen analysis
of $\overline{p}p$ scattering data.

\nonum\subsection{\bf II. STATISTICAL CONSIDERATIONS}
The Nijmegen database of $pp$ scattering data below
$T_{\rm lab}$ = 350 MeV contains at present $N_{\rm obs}$ = 1656
scattering observables. Since there are 119 experimental groups
with a finite overall normalization error as well as 12 additional
normalization parameters (from angle-dependent normalizations)
we have a total of $N_{\rm dat}$ = 1787 $pp$ data.
In the Nijmegen partial-wave analysis these
data are fitted with $N_{\rm par}$ = 22 model parameters,
which includes the $pp\pi^0$ coupling constant.
Because there are also 22 groups with a floated normalization
the number of degrees of freedom is $N_{\rm df}$ = 1612.
If the database is a correct statistical ensemble
and if the theoretical model is correct, one
expects $\langle\chi^2_{\rm min}\rangle$ =
$N_{\rm df}$ $\pm$ $\sqrt{2N_{\rm df}}$ = 1612 $\pm$ 57.
In our latest analysis we reach $\chi^2_{\rm min}$ = 1786,
which is only 174, or 3 standard deviations, higher than the
expectation value. This difference is at least partially due
to small theoretical shortcomings in our model.
This implies, therefore, that there is still some room for
theoretical improvements in the $pp$ partial-wave analysis.
We have investigated the statistical quality of the
final $pp$ data set by calculating the momenta of the
theoretically expected $\chi^2$-distribution and comparing them
to the ones actually found in the analysis.
The results are presented in Table~\ref{tab:mom}.
For details about the statistical tools used in the Nijmegen
partial-wave analyses we refer to Ref.~\cite{Ber88}.
The second and higher central moments found in the analysis are
in excellent agreement with their expectation values. This shows
that the statistical quality of the $pp$ data set used is
very good. It shows that our analysis with its
$\chi^2_{\rm min}/N_{\rm df}$ = 1.108 is already quite good.
Nevertheless, we hope that a significant drop in $\chi^2_{\rm min}$
can still be obtained giving better agreement between $N_{\rm df}$
and $\chi^2_{\rm min}$. In this way our partial-wave analysis
becomes a tool, because it can decide on the quality of the
proposed improvements in theory.

We have demonstrated that our multienergy partial-wave solution
is essentially correct statistically. This has rather strong
consequences: it means that the values for the phase shifts and
coupling constants as well as the statistical errors on these
quantities as determined in the multienergy analyses are essentially
correct. Here the statistical error on a particular quantity
(e.g., a phase shift) is the error as obtained in the standard way
via the $\chi^2$-rise-by-one rule, as discussed for example in
Sec.\ V\,A\,2 of Ref.~\cite{Ber88}.
Including new experiments in the multienergy analysis will change
the phase shifts and coupling constants not more than
1 or 2 multienergy standard deviations. The same is not necessarily
true for quantities determined in a single-energy analysis.
For example, in a single-energy $np$ analysis around 100 MeV
there are no $np$ spin-correlation data to pin down the $
\varepsilon_1$ mixing parameter, whereas in the multienergy analysis
the presence of spin-correlation data at the adjoining energies near
50 and 150 MeV also allows for an accurate determination of
the $\varepsilon_1$ mixing parameter at 100 MeV.
Therefore, the multienergy values and errors for the
phase shifts and coupling constants are much more realistic than
the single-energy determinations.

In our latest $pp$ analysis~\cite{Elb91} we used
a parametrization for the $^1S_0$ partial wave different from
the one used in previous analyses~\cite{Ber90,Sha90}.
This parametrization has less
parameters but gives a somewhat higher $\chi^2_{\rm min}$.
We feel, however, that it is a better parametrization,
since the $^1S_0$ wave was probably over-parametrized in
the older analyses. This means that results presented for
this analysis are not necessarily also true for the older
analyses, especially with respect to statements about the
$^1S_0$ wave. In spite of a large amount of effort on
our side, the description of this partial wave is still
not as good as we would like.

In our $pp$ analysis we determine the $pp\pi^0$ coupling
constant at the pion pole. We find now $f^2_{pp\pi^0}$ = 0.0750(5),
where the error is purely statistical.
If we fix this coupling at the old value 0.079 the result is
$\chi^2_{\rm min}$ = 1842, which is $\Delta\chi^2_{\rm min}$ = 56,
or 7.5 standard deviations, higher than the minimum.
We stress again that no particular data set is responsible for
the specific value of the coupling constant, but that all
data when fitted in a multienergy partial-wave analysis
contribute to this value.
{}From these numbers one can get a feeling about
how unreliable it must be to make statements
on the pion-nucleon coupling constant with the help
of potential models which fit the data with
$\chi^2_{\rm min}/N_{\rm dat}$ $\simeq$ 2,
instead of drawing conclusions based on multienergy
analyses with $\chi^2_{\rm min}/N_{\rm dat}$ $\simeq$ 1.
Nevertheless, in some recent papers~\cite{Mac91,Hai92}
conclusions about the pion-nucleon coupling constant
are drawn with the help of potential models
that have $\chi^2_{\rm min}/N_{\rm dat}$ $\simeq$ 2 or
$\chi^2_{\rm min}$ $\simeq$ 3600 which is about 2000,
or 35 standard deviations, and not just 174, or 3 standard
deviations, higher than the expectation
value $\langle\chi^2_{\rm min}\rangle$ = 1612.
The point we want to make here is that
these potential models can still be improved
in so many different places in so many different ways,
that it is very presumptuous to try to make any conclusions
about a $\Delta\chi^2_{\rm min}$ = 56 effect. All such
conclusions are inevitably very model dependent, which seems
to be generally overlooked by the authors criticizing us.

\nonum\subsection{\bf III. FORM FACTORS}
Especially persistent is the suggestion of systematic errors due
to form-factor effects~\cite{Tho89,Gro90,Hol90,Coo90,Eri92,Hai92},
although we have repeatedly~\cite{Klo91,Tim91,Sha90,Elb91}
stressed that we determine the pion-nucleon coupling constant
at the pion pole, and that therefore
form factors are irrelevant to the issue.
We did explicitly check this by adding an exponential form
factor to the pion-exchange potential in our analysis. We used
\begin{equation}
   F({\bf k}^2) = \exp[-({\bf k}^2+m_{\pi}^2)/\Lambda_0^2]  \:\: ,
\end{equation}
normalized such that at the pion pole $F(-m_{\pi}^2)=1$.
For values of the cutoff mass $\Lambda_0$ as low as 500 MeV
we have found no significant changes. This can be seen in
Table~\ref{tab:form}, where the results for the $pp\pi^0$ coupling
constant and the corresponding values for $\chi^2_{\rm min}$
are presented as a function of the cutoff mass $\Lambda_0$ in MeV.
Evidently there are no significant changes for realistic
values of the cutoff mass. In spite of our statements, however,
it seems to have been suggested recently in the panel discussion
at the Adelaide conference~\cite{Eri92} that the value of the
coupling constant may depend critically on the shape
of the form factor and that we, like other groups,
should use a form factor of the Feynman type
\begin{equation}
   F({\bf k}^2)= \left[ \frac{\Lambda_2^2-m_{\pi}^2}
                             {\Lambda_2^2+{\bf k}^2} \right]^2 \:\: ,
\end{equation}
again normalized such that at the pion pole $F(-m_{\pi}^2)=1$.
The square appears because one takes a
monopole form factor at each vertex.
Let us start by saying that we strongly feel that an exponential
form factor is more physical than a Feynman-type form factor.
The exponential form factor
follows quite naturally from Regge-pole theory and from
constituent-quark models with harmonic-oscillator wave functions.
A Feynman-type form factor, on the other hand,
which is essentially a phenomenological
regulator, has an annoying singularity at ${\bf k}^2$
= $-\Lambda_2^2$ in the unphysical region. This Feynman-type
form factor is chosen mainly for reasons of convenience.

To our mind, it is hard to understand how the {\bf type} of form
factor can matter once it has been demonstrated, using one
particular type, that the value of the coupling constant
is determined at the pole. When comparing form factors,
it should always be kept in mind that the
form factor is related to the size of the nucleon.
An exponential form factor and a Feynman-type form factor
give approximately the same nucleon size when
$\Lambda_0$ = $\Lambda_2/\sqrt{2}$. Reliable values of
the cutoff mass are hard to determine in $N\!N$ scattering
since these depend, for instance, on which heavy mesons are
included in the model~\cite{Hai92}.
But, in general, if a cutoff mass is used which affects the
pion-exchange potential drastically at distances where the
nucleons do not overlap anymore, the obvious conclusion should
be that this is an unrealistic low value for the cutoff mass.
As an example, look at Figure 4 in Ref.~\cite{Eri92}, where one
is willing to accept a 20\% reduction of the tensor force
due to pion exchange at a distance of 2.5 fm.
In the Nijmegen soft-core potential~\cite{Nag78} the
exponential cutoff mass is 965 MeV and in the Bonn
potential~\cite{Mac87} the Feynman-type cutoff mass for the
pion-exchange potential is 1300 MeV. In fact, it has,
until very recently~\cite{Hai92}, always been claimed by the
Bonn group that a satisfactory fit to the $N\!N$ scattering
data is impossible with a lower cutoff mass.
If one looks at potential models, a value of 500 MeV
for a cutoff mass in a form factor is quite low.

To meet all criticism, however, and to avoid new
misconceptions, we have again explicitly checked our
conjectures by repeating the $pp$ analysis using a
pion-exchange potential with Feynman-type form factor.
As we expected, our findings were entirely similar to
those with an exponential form factor.
We conclude therefore that neither the shape of the form factor
nor the value of its cutoff mass (as long as it is not
unreasonably low) has a significant influence on our determination
of the pion-nucleon coupling constant. The obvious reason for
this nice feature is to be found in the specific method
of analysis which allows the extraction of the coupling constant
from the asymptotic behavior of the one-pion-exchange potential
in configuration space and its determination is not
sensitive to short-range modifications.

\nonum\subsection{\bf IV. DETERMINATION FROM \\
                          DIFFERENT OBSERVABLES, \\
                          PARTIAL WAVES, AND ENERGY RANGES}
It is an interesting exercise to investigate
which particular types of observables are the
most sensitive to variations in the coupling constant.
We have repeated the $pp$ analysis for 4
different values of the $pp\pi^0$ coupling constant and the $np$
analysis, with the $N\!N\pi^0$ coupling constants fixed at 0.075,
for 4 values of the $np\pi^{\pm}$ coupling constant.
The resulting values of $\chi^2_{\rm min}$ are tabulated in
Table~\ref{tab:f2pp} for the different types of $pp$ scattering
observables and in Table~\ref{tab:f2np} for the $np$ observables.
The value found for the charged-pion coupling constant in
the $np$ analysis is $f^2_{\mbox{\scriptsize c}}$ = 0.0748(3).
It can be seen that in the $np$ analysis the difference between
$f^2_{\mbox{\scriptsize c}}$ =
0.075 and 0.079 is $\Delta\chi^2_{\rm min}$ = 255, so this
is a difference of 16 standard deviations! We did already mention
above that the corresponding difference in the $pp$ analysis,
for the $pp\pi^0$ coupling constant, amounts to
$\Delta\chi^2_{\rm min}$ = 56, or 7.5 standard deviations.
Furthermore, it can be seen from these Tables that no
particular type of observable is solely responsible for
the low value of the coupling constant, but that essentially
all types of $pp$ as well as $np$ scattering data favor a low
pion-nucleon coupling constant.

In order to investigate in what way the different partial
waves are sensitive to the $pp\pi^0$ coupling constant,
we introduced first of all two different couplings,
one for the singlet waves and one for the triplet waves.
We then find $\chi^2_{\rm min}$ = 1786 for 1611 degrees of freedom.
The resulting coupling constants are 0.0753(7) for the singlet
waves and 0.0750(6) for the triplet waves. This shows that
both singlet and triplet waves favor a low coupling constant.

Next, we introduced in turn a different $pp\pi^0$
coupling constant for a specific partial wave and
one for all other partial waves. This exercise was done for
all parametrized waves in the analysis: $^1S_0$, $^1D_2$,
$^1G_4$, $^3P_0$, $^3P_1$, $^3P_2$--$^3F_2$, $^3F_3$, and
$^3F_4$--$^3H_4$. The results for the
coupling constants are shown in Table~\ref{tab:f2wave}.
For all cases excepting the $^1S_0$ wave
the results favor a low coupling constant.
In the case that a separate $pp\pi^0$ coupling
was introduced in the $^1S_0$ channel
an improvement of $\Delta\chi^2_{\rm min}$ = 6.6 was found for
a high coupling constant. This is presumably a consequence of
our new parametrization of the $^1S_0$ partial wave which
is evidently not perfect. When the coupling constant in the
$^1S_0$ wave is fixed at 0.075 the results for all remaining
partial waves are nicely consistent with a $pp\pi^0$ coupling
constant somewhat lower than 0.075. Apparently the $^1S_0$
wave enhances this to $f^2_{pp\pi^0}$ = 0.0750(5).
This is probably also the reason that of the different types
of observables only the differential cross sections favor
a coupling somewhat larger than 0.075 (see Table~\ref{tab:f2pp}),
since differential cross sections are more sensitive to
this wave than spin-dependent observables. There is thus some
indication for a small systematic error pertaining to
the $^1S_0$ partial wave in the $pp$ analysis.
Further study is required to find out to what extent this is
to be attributed to friction between the data or to a flaw
in our theoretical treatment of the $^1S_0$ channel.
In view of the above findings, it may be better to fix the
coupling constant in the $^1S_0$ channel at a value of 0.075
and determine $f^2_{pp\pi^0}$ from the remaining partial waves.
We then find for 1612 degrees of freedom $\chi^2_{\rm min}$ =
1787 and the coupling constant becomes $f^2_{pp\pi^0}$ = 0.0746(6).
This is within one standard deviation from the value quoted above
$f^2_{pp\pi^0}$ = 0.0750(5) determined from all partial waves
including the $^1S_0$.

We can also bypass the apparent friction in the $^1S_0$ partial
wave by removing from the database the data
taken at very low energies below 1 MeV, namely
the Los Alamos cross sections around the interference minimum
measured by Brolley {\it et al}.~\cite{Bro64} and the Z\"urich
cross sections from Thomann {\it et al}.~\cite{Tho78}.
The reason, of course, is that at these low energies the $^1S_0$
phase shift is very accurately known and gives a very strong
constraint on the parametrization of the $^1S_0$ phase shift.
The results from this 3-350 MeV partial-wave analysis are
presented in Table~\ref{tab:tlab}. As expected, we find a
somewhat smaller coupling constant $f^2_{pp\pi^0}$ = 0.0743(6)
instead of 0.0750(5).
If we fix the coupling constant in the $^1S_0$ channel at 0.075,
we find in the 3-350 MeV analysis $f^2_{pp\pi^0}$ = 0.0744(6).
So there are strong indications that the value for
the neutral-pion coupling constant as determined in the
$pp$ partial-wave analysis is somewhat smaller than 0.075.
In the last line of Table~\ref{tab:pion} we quote
$f^2_{pp\pi^0}$ = 0.0745(6).

In order to demonstrate that it is not only the data at low energies
that pin down the coupling constant, Table~\ref{tab:tlab} also
contains the results for the $pp\pi^0$
coupling constant obtained from a number of analyses in
different energy ranges. It can be seen from this Table
that the data at energies higher than 10 MeV or 30 MeV favor
a low coupling constant as well.
Similar results are found if we restrict the energy range at the
high end, by doing an analysis of the data up to, say, 280 MeV.
These findings once more underline our claim that the database as
a whole contributes to a low value for the pion-nucleon coupling
constant, and not some particular experiment(s) or the data
in a restricted energy bin.

\nonum\subsection{\bf V. \mbox{$pp$}
                         ANALYZING-POWER DATA \\ AROUND 10 MeV}
Let us next turn to our investigations of specific
experiments that are discussed in connection
with the pion-nucleon coupling constant.
In Ref.~\cite{Hai92}, for instance, it is stated
that the prime reason for a low $pp\pi^0$ coupling constant
was our analysis of $pp$ analyzing-power data around 10 MeV.
In Ref.~\cite{Eri92}, the same statement can be found in a
different form, where it is said that the $^3P$ phase shifts
around 10 MeV are very important in the determination.
However, these arguments only reflect our statements based on
a preliminary $pp$ analysis by our group~\cite{Ber87}.
Apparently, these critics have overlooked our amendment to these
statements as discussed in our paper on the completed 0-350 MeV
$pp$ analysis~\cite{Ber90}, where we incorporated many
theoretical improvements and included much more experimental data.
In this latter paper it is explicitly stated
that the $^3P$ waves are not especially
important in the determination of $f^2_{pp\pi^0}$
and that it is not possible to pinpoint
some specific type of observables as particularly constraining.
Concerning the analyzing-power data around 10 MeV,
the 15 Wisconsin data points at 9.85 MeV~\cite{Bar82} have in our
latest multienergy $pp$ analysis $\chi^2_{\rm min}$ = 16.
If we fix the $pp\pi^0$ coupling constant at 0.079 and refit,
$\chi^2_{\rm min}$ on these data increases to 31. If we leave out
this group (lowering the number of degrees of freedom with 15 to
1598), $\chi^2_{\rm min}$ drops about 16 from 1786 to 1770 and the
value for the coupling constant becomes $f^2_{pp\pi^0}$ =
0.0751(6). This clearly shows that these data are not alone
responsible for the low value of the coupling constant,
although this group clearly favors a low coupling constant.
We stress once more, however, that this latter conclusion
is only justified when reached in a multienergy partial-wave
analysis using the data as a whole.
We did already demonstrate that the data above 30 MeV give for the
coupling constant $f^2_{pp\pi^0}$ = 0.0743(6), so the analyzing-power
data around 10 MeV are absolutely not crucial to a low value for
$f^2_{pp\pi^0}$.

A criticism of our $pp$ analysis in this context is the fact that
we do not include in our database another group of analyzing-power
data around 10 MeV (and 25 MeV) taken by the Erlangen group
of Kretschmer {\it et al}.~\cite{Kre86,Kre89}. We do not
do this because it is our policy not to include data
that have not been published in a regular physics journal.
Moreover, in this specific case we have committed ourselves
to not publishing any analysis of these specific data prior
to their publication by the Erlangen group. Of course we are well
aware of the existence of these data and we did analyze them.
We can state that we find no reason whatsoever to modify
any of our conclusions regarding the $pp\pi^0$ coupling
constant. It is definitely not true that it is crucial which
one of these two data sets (the Wisconsin or the Erlangen set)
around 10 MeV is included, as is concluded in
Ref.~\cite{Hai92} from a study with meson-exchange potential models
that have $\chi^2_{\rm min}/N_{\rm dat}$ $\agt$ 2.

\nonum\subsection{\bf VI. \mbox{$np$}
                        BACKWARD DIFFERENTIAL \\ CROSS SECTIONS}
A point of discussion regarding the $np$ analysis is the
normalizations of the $np$ differential cross sections and
their relation to our determination of the $np\pi^{\pm}$
coupling constant (see, for instance, Ref.~\cite{Eri92}).
It is common folklore that the $np\pi^{\pm}$ coupling constant
is determined mainly by the peak present in backward $np$
differential cross sections. It was suggested by Chew~\cite{Che58}
as early as 1958 that this is a good place to extract
the pion-nucleon coupling constant. If this is true, then a very
important group of data should be the Los Alamos set of backward
cross sections measured by Bonner {\it et al}.~\cite{Bon78}.
However, we have seen already from Table~\ref{tab:f2np} that
all observables, and not the differential cross sections in
particular, favor a low coupling constant.
Our results and conclusions regarding the data from Bonner
{\it et al}. are the following.
The way we handle the normalization of a group of data
and its uncertainty is explained in detail in Ref.~\cite{Ber88}.
One group of 42 cross sections at 194.5 MeV is rejected completely,
as well as 1 data point at 344.3 MeV.
For the remaining 607 backward $np$ cross sections
at 10 different energies we find $\chi^2_{\rm min}$ = 630 for
$f^2_{\mbox{\scriptsize c}}$ = 0.075 and $\chi^2_{\rm min}$ = 655
for $f^2_{\mbox{\scriptsize c}}$ = 0.079, so
$\Delta\chi^2_{\rm min}$ = 25. Of these 10 groups,
7 groups have a floated normalization and 3 groups have a finite
normalization error of 4\%, where the sensitivity of these last 3
groups (242 data points) to $f^2_{\mbox{\scriptsize c}}$ is
rather small.
So we see that it is mainly the {\bf shape} of the cross section
and not the normalization that makes that these Los Alamos data
favor a low coupling constant. In Table~\ref{tab:bonner} we give
the results obtained for the charged-pion coupling constants for
the 10 individual groups by fitting a parabola through 4 values
of $\chi^2_{\rm min}$ for 4 different coupling constants.
We see that most groups favor a low coupling constant, but
the errors are rather large.

We have also tabulated the norm of these groups as determined in the
multienergy analysis, once again following the $\chi^2$-rise-by-one
rule using the full error matrix. It can be seen that our multienergy
solution pins down these normalizations with very small errors, of the
order of $0.5\%$, which is much smaller than the $4\%$ error quoted by
the experimentalists. There is essentially no difference
between the groups with a floated normalization and the groups
with a finite normalization error.

Our conclusion is that the relevance of the backward cross
sections for determining the charged-pion coupling constant
is more limited than is generally assumed.
There are other experiments that are much more constraining
for the coupling constant. Here we mention the 12 analyzing-power
data at 10.03 MeV taken by Holslin {\it et al}.~\cite{Hol88},
the 16 spin correlations measured by Bandyopadhyay
{\it et al}.~\cite{Ban89} at 220 MeV, and the 19
spin correlations taken by the same group at 325 MeV.
Again, these statements apply to an analysis of the groups
within multienergy partial-wave analyses of the complete database,
and do not follow from studies of the individual experiments.
However, we want to stress once more that our low value of the
charged-pion coupling constant is not only due to these accurate
analyzing-power and spin-correlation data. The analysis without
these data still yields $f^2_{\mbox{\scriptsize c}}$ = 0.0750(4),
demonstrating that also the other data favor a low, but
slightly less accurate, value.

\nonum\subsection{\bf VII. DETERMINATION FROM \\ CHARGE-EXCHANGE DATA}
In this section, we add some remarks about our coupled-channels
partial-wave analysis~\cite{Tim91} of
antiproton scattering data. In this case a neutral pion can
be exchanged in elastic $\overline{p}p \rightarrow \overline{p}p$
scattering and a charged pion in charge-exchange
$\overline{p}p \rightarrow \overline{n}n$ scattering.
Until recently it was believed by probably everybody
(including ourselves) that a partial-wave analysis of
these reactions was out of the question.
We find it gratifying that the methods used in the partial-wave
analyses of $pp$ and $np$ scattering data could be extended to
the case of the antiproton elastic and charge-exchange scattering.
The value for the $np\pi^{\pm}$ coupling
constant $f^2_{\mbox{\scriptsize c}}$ = 0.0751(17)
found in our analyses of charge-exchange data is in nice
agreement with the values found in the analyses of $N\!N$ data.
In fact, if it is possible to measure the differential cross
section for $\overline{p}p \rightarrow \overline{n}n$
with the accuracy stated by Bradamante (private communication,
see also Ref.~\cite{Mac92}), the charge-exchange
reaction will be an even more competitive place to study
the isovector-meson coupling constants.

We want to stress that the still popular (but now rather outdated)
few-parameter optical-potential models can in no way be compared
to a sophisticated partial-wave analysis. In the first approach
at best a crude qualitative description of a limited number of
data is possible. No $\chi^2_{\rm min}$ is ever presented.
It is easy for us to construct a similar optical-potential
model, by supplementing the C-parity-transformed Nijmegen
potential~\cite{Nag78} by an imaginary potential containing 2 free
parameters. We then find at best $\chi^2_{\rm min}$ $\sim$ $10^9$
for a database of 3309 observables. This should be compared to
$\chi^2_{\rm min}$ = 3592.5 reached in our multienergy partial-wave
analysis on the same set of data.
Of course, our two-parameter model is as bad or as good as
any other few-parameter optical-potential model. For instance,
for the 1968 prototype Bryan-Phillips model~\cite{Bry68}
we find a $\chi^2_{\rm min}$ which is even much larger.
One can never hope to describe all $\overline{p}p$ scattering
data with just 2 or 3 free parameters, when one needs already
about 20 free parameters to fit the $pp$ data. In a single-energy
$\overline{p}p$ analysis in principle 8 times as many phase-shift
parameters are required compared to a $pp$ analysis.
One should keep in mind that in the past these optical-potential
models were never intended for a quantitative comparison
to the data. The 1980 Dover-Richard model~\cite{Dov80},
for instance, served an excellent purpose in examining what could be
expected qualitatively when LEAR would come into operation in 1983.
At present, however, this approach seems hardly justified anymore.
In our opinion, these naive models that
do not fit the presently available data at all
are completely inadequate to address in a reliable manner
issues like the value of the pion-nucleon coupling constant, as was
attempted very recently in Ref.~\cite{Hai92}. This can be seen,
for instance, from the bad fit in Ref.~\cite{Hai92} to very recent
accurate charge-exchange analyzing-power data from
LEAR~\cite{Bir90}. After almost 10 years of data-taking at LEAR,
it unfortunately still is a common practice to compare
the data to the ``predictions'' of these museum models and then
draw strong conclusions about the physics behind these
models from such a comparison.

\nonum\subsection{\bf VIII. CONCLUSIONS}
To summarize, we firmly believe that the value of the
pion-nucleon constant found in the Nijmegen partial-wave
analyses of $pp$, $np$, and $\overline{p}p$ scattering data
is essentially correct and free of significant systematic
errors. An excellent $\chi^2_{\rm min}$ is reached in all cases,
reflecting both the statistical consistency of the data sets and
the quality of the analyses. The specific method of analysis
allows the extraction of the coupling
constant at the pion pole from the asymptotic pion-exchange
potential and ensures a clean separation from short-range
form-factor effects and heavy- or multi-meson-exchange forces.
We stress that in all cases we have also determined
the mass of the exchanged pion and always found
agreement with the experimental values. For instance, in the
combined analysis of $pp$ and $np$ data~\cite{Klo91} it was
found that $m_{\pi^0}$ = 135.6(1.3) MeV and $m_{\pi^{\pm}}$
= 139.4(1.0) MeV. This success is a very strong argument
against the presence of significant systematic errors,
such as form-factor effects. Furthermore, there is consistency
in the results from different analyses as well as agreement with
the value $f^2_{\mbox{\scriptsize c}}$ = 0.0735(15)
found by Arndt and coworkers in their latest VPI\&SU
analysis of $\pi^{\pm}p$ scattering data~\cite{Arn90,Arn91a,Arn91b}.

We pointed out before~\cite{Tim91} that the Goldberger-Treiman
relation~\cite{Gol58} also favors a low pion-nucleon coupling
constant. Our present results indicate a need for reconsideration
of calculations on the so-called Goldberger-Treiman discrepancy
(see, e.g., Ref.~\cite{Coo90}).
Recently, Workman, Arndt, and Pavan~\cite{Wor92}
showed that the Goldberger-Miyazawa-Oehme sum rule~\cite{Gol55}
provides rather model-independent evidence for a low coupling
constant as well.

In a recent study of the deuteron properties~\cite{Mac91}
it was shown that a low value for the pion-nucleon
coupling constant implies that the value for
$\kappa_{\rho}$ $\equiv$ $f_{N\!N\rho}/g_{N\!N\rho}$ = 6.6
as determined in the Karlsruhe-Helsinki analyses of $\pi{^\pm}p$
scattering data~\cite{Hoh75} must be wrong. It was further shown
that the preferred value for $\kappa_{\rho}$ is in
agreement with the value $\kappa_{\rho}$ = 4.2 as found in
the 1978 Nijmegen soft-core nucleon-nucleon potential~\cite{Nag78}
and with the value $\kappa_{\rho}$ = 3.7 which follows from
vector-meson dominance of nucleon electromagnetic form factors.

We feel that especially the value for the $pp\pi^0$ coupling
constant as determined in the $pp$ analysis is compelling
evidence for a low pion-nucleon coupling constant.
With the exception of the $^1S_0$ wave, the $pp\pi^0$
coupling constant can be determined from all parametrized
partial waves, all values being consistent.
Given the $pp\pi^0$ coupling constant, it seems to us
that claims for a high $np\pi^{\pm}$ coupling constant are
untenable, since in that case not only three independent recent
determinations of this coupling constant must be wrong, but
one also has to cope with a large breaking of charge-independence,
which is theoretically very difficult to accommodate~\cite{Tim91}.
Therefore, we strongly recommend that in future work on
nucleon-nucleon scattering the value $f^2_{N\!N\pi}$ = 0.0745
at the pion pole is taken as a starting point from which further
consequences can be discussed. \\ \\

\nonum\subsection{\bf ACKNOWLEDGMENTS}
\noindent
Discussions with Prof. D. Bugg, Dr. Th. Rijken, and
M. Rentmeester are gratefully acknowledged.
Part of this work was included in the research program of the
Stichting voor Fundamenteel Onderzoek der Materie (FOM) with
financial support from the Nederlandse Organisatie voor
Wetenschappelijk Onderzoek (NWO).

\newpage
\mediumtext
\begin{table}
\caption{Recent determinations of the pion-nucleon coupling constants
         from dispersion-relation (DR) analyses of pion-nucleon
         scattering data and from partial-wave analyses (PWA) of
         nucleon-nucleon and antinucleon-nucleon scattering data.}
\begin{tabular}{ccccc}
    Group        & Year &   Method            &
  $10^3\,f^2_{pp\pi^0}$  & $10^3\,f^2_{\mbox{\scriptsize c}}$ \\
\tableline
 Karlsruhe-Helsinki~\cite{Dum83} & pre-1983 & $\pi^{\pm}p$ DR &
                                       &   79(1)                     \\
 Nijmegen~\cite{Ber90}           & 1987-1990 &    $pp$ PWA    &
                           74.9(0.7)   &                             \\
 VPI\&SU~\cite{Arn90}            & 1990 & $\pi^{\pm}p$ DR     &
                                       &   73.5(1.5)                 \\
 Nijmegen~\cite{Klo91}           & 1991 & combined $N\!N$ PWA &
                           75.1(0.6)   &   74.1(0.5)                 \\
 Nijmegen~\cite{Tim91}           & 1991 & $\overline{p}p$ PWA &
                                       &   75.1(1.7)                 \\
 this work                       & 1992 & $pp$ and $np$ PWA   &
                           74.5(0.6)   &   74.8(0.3)                 \\
\end{tabular}
\label{tab:pion}
\end{table}

\narrowtext
\setdec 00.000
\begin{table}
\caption{Comparison between the moments of the $\chi^2$-probability
         distribution expected from theory and those determined in
         our partial-wave analysis (PWA) of $pp$ data.
         Tabulated are $\langle\chi^2_{\rm min}\rangle/N_{\rm df}$
         and the central moments $\mu_n$ for $n=2,3,4$.}
\begin{tabular}{ccc}
                 &      theory           &     PWA   \\ \tableline
$\langle\chi^2_{\rm min}\rangle/N_{\rm df}$
                 &  \dec 1.000$\pm$0.035 &    \dec  1.108 \\
   $\mu_2$       &  \dec  1.81$\pm$0.12  &    \dec  1.83  \\
   $\mu_3$       &  \dec  5.55$\pm$0.74  &    \dec  5.40  \\
   $\mu_4$       &  \dec 29.8$\pm$4.5    &    \dec 27.6   \\
\end{tabular}
\label{tab:mom}
\end{table}

\mediumtext
\begin{table}
\caption{The $pp\pi^0$ coupling constant as a function of
         the cutoff mass in the exponential form factor.
         The number of degrees of freedom is 1612.}
\begin{tabular}{cccccc}
$\Lambda_0$ (MeV) & 500.0   & 750.0   &1000.0   &1250.0   &$\infty$  \\
\tableline
$10^3\,f_{pp\pi^0}^2$
                  &75.2(0.5)&75.0(0.5)&75.0(0.5)&75.0(0.5)&75.0(0.5) \\
$\chi^2_{\rm min}$& 1786.3  & 1786.4  &1786.4   &1786.4   & 1786.4   \\
\end{tabular}
\label{tab:form}
\end{table}

\mediumtext
\begin{table}
\caption{$\chi^2$ results for 4 values of the $pp\pi^0$
         coupling constant for the different types
         of observables in the analysis of $pp$ scattering data.
         The numbers in the two last
         columns are obtained by fitting a parabola to the
         numbers in the four preceding columns. The number
         of degrees of freedom is 1613 for each value of
         $f^2_{pp\pi^0}$.}
\begin{tabular}{crrrrrrc}
  type     & $N_{\rm dat}$ &
    $10^3\,f^2_{pp\pi^0}$ = 73  &  75  &  77  & 79  &
  $\chi^2$(min)     &  $10^3\,f^2_{pp\pi^0}$(min) \\ \tableline
d$\sigma$/d$\Omega$ &
     821   &  838.7 &  825.7 &  823.0 &  830.8 &  822.7 & 76.5(0.9)  \\
  $A_y$             &
     558   &  585.2 &  580.0 &  585.7 &  602.2 &  580.0 & 75.0(0.9)  \\
 $A_{ii}$,$C_{nn}$  &
      66   &   52.8 &   55.5 &   60.0 &   66.3 &   51.9 & 71.0(2.1)  \\
 $D$,$D_t$          &
      97   &  105.2 &  107.5 &  112.0 &  118.9 &  104.9 & 72.0(1.9)  \\
 $R$,$R'$,$A$,$A'$  &
     209   &  194.5 &  193.1 &  194.9 &  199.8 &  193.1 & 74.9(1.6)  \\
  rest              &
      36   &   24.8 &   24.7 &   24.6 &   24.5 &        &            \\
 \tableline
  all               &
    1787   & 1801.2 & 1786.4 & 1800.2 & 1842.4 & 1786.4 & 75.0(0.5)  \\
\end{tabular}
\label{tab:f2pp}
\end{table}

\mediumtext
\begin{table}
\caption{$\chi^2$ results for 4 values of the $np\pi^{\pm}$
         coupling constant for the different types
         of observables in the analysis of $np$ scattering data.
         The numbers in the two last
         columns are obtained by fitting a parabola to the
         numbers in the four preceding columns. The $N\!N\pi^0$
         coupling constants are taken to be 0.075. The number
         of degrees of freedom is 2331 for each value of
         $f^2_{\mbox{\scriptsize c}}$.}
\begin{tabular}{crrrrrrc}
  type     & $N_{\rm dat}$ &
  $10^3\,f^2_{\mbox{\scriptsize c}}$ = 73  &  75  &  77  & 79  &
  $\chi^2$(min)  &  $10^3\,f^2_{\mbox{\scriptsize c}}$(min)
  \\ \tableline
 $\sigma_{\rm tot}$,$\Delta\sigma_{\rm L}$,$\Delta\sigma_{\rm T}$ &
    252 &  232.9 &  229.7 &  232.4 &  242.4 &  229.5 & 75.1(1.1)  \\
d$\sigma$/d$\Omega$           &
   1350 & 1379.0 & 1364.2 & 1367.9 & 1391.8 & 1363.2 & 75.6(0.6)  \\
  $A_y$                       &
    738 &  737.7 &  720.3 &  745.9 &  830.4 &  717.8 & 74.8(0.4)  \\
 $A_{yy}$,$A_{zz}$            &
     86 &   77.2 &   72.6 &   91.2 &  136.0 &   71.2 & 74.4(0.6)  \\
 $D_t$                        &
     43 &   42.8 &   39.8 &   42.0 &   51.6 &   39.5 & 75.1(1.1)  \\
 $R_t$,$R'_t$,$A_t$,$A'_t$    &
     43 &   54.6 &   58.6 &   68.5 &   88.4 &   54.7 & 73.1(1.0) \\
\tableline
  all                         &
   2512 & 2524.3 & 2485.3 & 2547.9 & 2740.7 & 2480.4 & 74.8(0.3)  \\
\end{tabular}
\label{tab:f2np}
\end{table}

\narrowtext
\begin{table}
\caption{Results for the pion-nucleon coupling constant
         introduced separately in each parametrized partial
         wave in the analysis of $pp$ scattering data.
         For fitting the coupling constant in non-S waves
         the coupling in the $^1S_0$ wave is fixed at 0.075.
         The number of degrees of freedom is 1611 in each case.}
\begin{tabular}{cccc}
 partial wave     &
  $10^3\,f^2$(wave) & $10^3\,f^2$(rest) & $\chi^2_{\rm min}$ \\
 \tableline
  $^1S_0$         &  79.7(1.9)  &  74.5(0.6)  &   1779.8     \\
  $^1D_2$         &  74.6(0.8)  &  74.6(0.6)  &   1786.0     \\
  $^1G_4$         &  74.6(2.1)  &  74.6(0.6)  &   1786.0     \\
  $^3P_0$         &  72.7(1.7)  &  74.8(0.6)  &   1784.6     \\
  $^3P_1$         &  74.9(0.7)  &  74.3(0.8)  &   1785.6     \\
 $^3P_2$--$^3F_2$ &  74.7(0.8)  &  74.6(0.6)  &   1786.0     \\
  $^3F_3$         &  73.3(1.3)  &  74.8(0.6)  &   1784.8     \\
 $^3F_4$--$^3H_4$ &  75.1(0.9)  &  74.5(0.6)  &   1785.6     \\
\end{tabular}
\label{tab:f2wave}
\end{table}

\narrowtext
\begin{table}
\caption{Values for the $pp\pi^0$ coupling constant
         determined in a partial-wave analysis
         $pp$ scattering data within different
         energy ranges in MeV.}
\begin{tabular}{crlcc}
  $T_{\rm lab}$ range & $N_{\rm df}$ &
  $10^3\,f^2_{pp\pi^0}$ & $\chi^2_{\rm min}$  &
  $\chi^2_{\rm min}/N_{\rm df}$ \\ \tableline
    0-350   &   1612   &   75.0(0.5)   &   1786.4  &  1.108  \\
    3-350   &   1435   &   74.3(0.6)   &   1596.6  &  1.113  \\
   10-350   &   1312   &   73.7(0.7)   &   1488.8  &  1.135  \\
   30-350   &   1237   &   74.2(0.8)   &   1397.6  &  1.130  \\
    0-280   &   1243   &   75.5(0.6)   &   1389.3  &  1.118  \\
    3-280   &   1066   &   74.5(0.7)   &   1189.7  &  1.116  \\
\end{tabular}
\label{tab:tlab}
\end{table}

\narrowtext
\begin{table}
\caption{The charged-pion coupling constant
         $f^2_{\mbox{\scriptsize c}}$ determined from
         the backward $np$ cross sections of Bonner
         {\it et al}.~\cite{Bon78} in the $np$
         partial-wave analysis. The numbers are
         obtained by fitting a parabola through
         the $\chi^2_{\rm min}$ results for 4 different
         coupling constants. For $T_{\rm lab}$ = 265.8 MeV
         these 4 numbers were consistent with a straight
         line.}
\begin{tabular}{crccc}
  $T_{\rm lab}$ (MeV) & $N_{\rm dat}$ &
  $\chi^2$(min) &  norm  &  $10^3\,f^2_{\mbox{\scriptsize c}}$(min)
  \\ \tableline
    162.0    &     43   &   60.0  &  1.092(7)  &   69.9(3.0)  \\
    177.9    &     44   &   44.0  &  1.083(7)  &   70.2(3.1)  \\
    211.5    &     43   &   31.0  &  1.063(7)  &   72.8(3.3)  \\
    229.1    &     49   &   62.3  &  1.058(7)  &   69.5(3.6)  \\
    247.2    &     53   &   38.5  &  1.042(7)  &   69.7(9.3)  \\
    265.8    &     63   &   ---   &  1.028(6)  &    ---       \\
    284.8    &     73   &   79.7  &  1.052(5)  &   75.3(3.5)  \\ \hline
    304.2    &     80   &   79.9  &  1.003(4)  &   74.6(3.4)  \\
    324.1    &     82   &   91.7  &  1.057(5)  &   78.0(4.3)  \\
    344.3    &     80   &   74.6  &  1.035(5)  &   74.8(3.9)  \\
\end{tabular}
\label{tab:bonner}
\end{table}

\end{document}